# Title: Intrinsic Donor-Bound Excitons in Ultraclean Monolayer Semiconductors

**Authors:** Pasqual Rivera[1], Minhao He[1], Bumho Kim[2], Song Liu[2], Carmen Rubio-Verdú[3], Hyowon Moon[4], Lukas Mennel[4], Daniel A. Rhodes[2], Hongyi Yu[5], Takashi Taniguchi[6], Kenji Watanabe[7], Jiaqiang Yan[8,9], David G. Mandrus[8-10], Hanan Dery[11], Abhay Pasupathy[3], Dirk Englund[4], James Hone[2#], Wang Yao[5#], & Xiaodong Xu[1,12#]

**Affiliations:**

[1]Department of Physics, University of Washington, Seattle, Washington 98195, USA

[2]Department of Mechanical Engineering, Columbia University, New York, NY, 10027, USA

[3]Department of Physics, Columbia University, New York, NY, 10027, USA

[4]Department of Electrical Engineering and Computer Science, Massachusetts Institute of Technology, Cambridge, MA, 02139, USA

[5]Department of Physics, University of Hong Kong, and HKU-UCAS Joint Institute of Theoretical and Computational Physics at Hong Kong, China

[6]International Center for Materials Nanoarchitectonics, National Institute for Materials Science, Tsukuba, Ibaraki 305-0044, Japan.

[7]Research Center for Functional Materials, National Institute for Materials Science, Tsukuba, Ibaraki 305-0044, Japan.

[8]Materials Science and Technology Division, Oak Ridge National Laboratory, Oak Ridge, Tennessee, 37831, USA

[9]Department of Materials Science and Engineering, University of Tennessee, Knoxville, Tennessee, 37996, USA

[10]Department of Physics and Astronomy, University of Tennessee, Knoxville, Tennessee, 37996, USA

[11]Department of Electrical and Computer Engineering, University of Rochester, Rochester, New York, 14627, USA

[12]Department of Materials Science and Engineering, University of Washington, Seattle, Washington, 98195, USA

[#]Correspondence to xuxd@uw.edu; wangyao@hku.hk; jh2228@columbia.edu

**Abstract**: **The monolayer transition metal dichalcogenides are an emergent semiconductor platform exhibiting rich excitonic physics with coupled spin-valley degree of freedom and optical addressability. Here, we report a new series of low energy excitonic emission lines in the photoluminescence spectrum of ultraclean monolayer $WSe_2$. These excitonic satellites are composed of three major peaks with energy separations matching known phonons, and appear only with electron doping. They possess homogenous spatial and spectral distribution, strong power saturation, and anomalously long population (> 6 μs) and polarization lifetimes (> 100 ns). Resonant excitation of the free inter- and intra-valley bright trions leads to opposite optical orientation of the satellites, while excitation of the free dark trion resonance suppresses the satellites photoluminescence. Defect-controlled crystal synthesis and scanning tunneling microscopy measurements provide corroboration that these features are dark excitons bound to dilute donors, along with associated phonon replicas. Our work opens opportunities to engineer homogenous single emitters and explore collective quantum optical phenomena using intrinsic donor-bound excitons in ultraclean 2D semiconductors.**

## Introduction

A promising route for optical encoding of matter is to employ excitons, Coulomb-bound electron-hole pairs, which are elementary optical excitations in semiconductors. Monolayer transition metal dichalcogenides (TMDs) are an emergent platform for exploring excitonic physics at the two-dimensional (2D) limit. This is largely due to their strong light-matter interactions, easy access to electric and magnetic control, and unique combination of spin-valley coupling and valley contrasting circular dichroism[1-4]. Encapsulation of monolayer TMDs within hexagonal boron nitride (hBN) has led to drastically improved sample quality[5,6], allowing the identification and detailed studies of a variety of optically bright and dark valley excitonic states[7-21], and their phonon replicas[22,23] assisted by both zone center[24,25] and zone edge phonons[26,27].

Despite the rapid progress in sample quality and understanding of their excitonic physics, these monolayer semiconductors are still far from perfect, and questions remain. For example, while localized single photon emitters have been observed in these materials[28-31], they exhibit random emission energies over a broad spectral range (>100 meV), and commonly lose the desirable valley optical selection rules due to the random anisotropy. Moreover, they generally appear in low quality samples, where inhomogeneous broadening obscures the underlying rich excitonic manifold that has been observed in clean samples. While the precise nature of these quantum light sources remains unclear, O interstitials[32,33] and extrinsic confinement potentials such as strain appear to play a role in the localized emission[34-38]. However, even the cleanest samples reported to date inevitably contain intrinsic defects (e.g. self-flux growth $WSe_2$ crystal with defect density of ~$1x10^{11}$ $cm^{-2}$)[39]. Several of these native defects have been identified and their electronic structure probed using scanning tunneling microscopy and spectroscopy[40-43]. A natural question arises as to if new spectral features may arise from excitons bound to such dilute intrinsic defects in these ultraclean monolayer crystals[44].

In this work, we report the observation of donor bound dark excitons and their phonon replicas in ultraclean monolayers of $WSe_2$ encapsulated in hBN. The samples in our study (N=10) were fabricated from different sources of $WSe_2$ and hBN bulk crystals and yielded consistent and reproducible results. The data presented in the main text are mainly from two devices. For electrostatic control of the charge carrier density in the $WSe_2$, a local graphite bottom gate is used. Figures 1a and 1b show an optical microscope image of Device 1 and its schematic, respectively.

The following experiments were performed at a temperature of either 1.6 or 4 K, and with excitation laser of energy 1.96 eV, unless otherwise specified.

## Results

### Observation of robust excitonic satellites

The PL intensity of Device 1 as a function of back gate voltage ($V_b$) and emission energy is shown in Figure 1c. The excitation power is 5 μW. All reported excitonic features such as excitons, trions, dark states, and various phonon replicas in the spectral range of 1. 65 to 1.75 eV are well-resolved (Supplementary Figure 1). The sample is also nearly intrinsic, with charge neutrality occurring at $V_b \approx$ -0.25 V. All these factors attest to the high sample quality. An observation is the emergence of low energy excitonic satellites occurring between 1.58 and 1.63 eV under n-type doping. The three main features, indicated as S1, S2, and S3 from high to low energy, are the focus of this paper.

The excitonic satellites have homogeneous spatial distribution and energy across the entire sample. Figures 1d displays the spatial map of the satellite emission intensity (spectral range shown on top of Fig. 1c at $V_b$ = 0.5 V), while Figure 1e shows the spatial distribution of S1 peak energy. Its peak position relative to neutral exciton is unchanged across the whole sample (See Supplementary Figure 2). Remarkably, the spectral structure and energetic positions of the excitonic satellites are also consistent across many samples. The normalized PL spectrum of three exemplary samples under similar experimental conditions is shown in Fig. 1f. The satellite emission energies, relative to the neutral exciton, between total 10 different samples fabricated over a 3-year span with different crystal sources are well aligned (Supplementary Figure 3). From the spatial and energetic homogeneity of the satellite PL, we can safely rule out their origin from contamination, cracks, or other extrinsic confinement potentials.

### Evidence of intrinsic defect bound excitons

The satellite emission exhibits distinct power dependence compared to the known free 2D excitons. Figure 2a presents the normalized PL spectra of Device 2 with powers ranging over three orders of magnitude (20 nW to 60 μW). The satellite emission dominates the entire spectrum at low powers, but quickly saturates and the free 2D excitonic manifold appears as the power increases. Above 10 μW excitation power, the satellite emission is overwhelmed by the linear response of the higher energy 2D exciton species (See Supplementary Figure 4 for additional samples). The strong power saturation is a hallmark of defect-localized excitons[45], and the low saturation threshold implies low defect density and long exciton lifetime. Note that the integrated photoluminescence saturation count rate is comparable to that of single photon emitters[31].

Circularly polarized optical pumping reveals non-trivial polarization of the satellite PL. We observe that the two highest-energy satellites, S1 and S2, are co-circularly polarized with the excitation laser, while the lowest energy satellite S3 is nearly unpolarized, as shown in Figure 2b for Device 2. The degree of circular polarization is defined as $\eta \equiv (I_+ - I_-)/(I_+ + I_-)$, where $I_\pm$ denotes the intensity of the $\sigma^\pm$ polarized components of the PL. For S1, we observe that $|\eta| \sim 0$ at low excitation power (< 100 nW), but increases with the excitation power to a value of $|\eta| \approx 0.6$ at powers above ~10 μW (inset of Figure 2b, integration over S1). Optical orientation of exciton spins is often observed in excitons bound to defects, where the localized electrons can become spin polarized, e.g. via efficient exchange interactions with photoexcited free electron spins[46]. The strong power dependence of circular polarization is thus another signature of localized excitonic spin states. We note that under linearly polarized excitation, the satellite emission is not polarized in the linear basis, regardless of the axis of excitation (Supplementary Figure 5). The lack of linear polarization and preservation of the circular optical selection rules imply underlying rotational symmetry ($C_3$) of the defect. This behavior is different from all reported quantum emitters found in $WSe_2$, which emit with linear polarization with strong spatial and spectral inhomogeneity[28-31].

Time resolved PL reveals anomalously long lifetime of the satellite states, as shown in Fig. 2c. A bi-exponential fit of the long PL component yields a population lifetime of $6.95 \pm 0.05$ μs. This is 3-5 orders of magnitude larger than that of both 2D bright and dark excitonic species[13,47,48], and consistent with the observed strong saturation of the satellite PL. We further extract the circular polarization lifetime of the satellite emission to be $116 \pm 4$ ns, as determined by fitting $\eta$ with a single exponential (inset of Fig. 2c). The polarization lifetime is also ~2-4 orders of magnitude longer than that of 2D bright[48] and dark excitons[41,42]. Such an enormously long population lifetime implies a dramatic reduction in the non-radiative lifetime of the satellite state. Moreover, the long polarization lifetime suggests small intervalley scattering rate, and weak transverse-longitudinal splitting, which further supports high symmetry ($C_3$) of the underlying confinement center.

All the above experimental features confirm that the excitonic satellites have distinct origin from both localized quantum emitters and free excitonic species previously reported in monolayer $WSe_2$. The spatial and spectral homogeneity across many samples, unusually long population and polarization lifetimes, and strong pump power dependence in both emission intensity and polarization, imply their origin as excitonic states bound to dilute intrinsic defects. The strong dependence of the observed PL features on electron doping is the first indication of the defect type as donors.

**Controllable synthesis of crystals with intrinsic defect bound excitons**

Figures 3a-c shows the wide field scanning tunneling microscopy (STM) topographic images of $WSe_2$ with three different growth parameters (see Methods, labeled as F1, F2, and F3). Two main types of defects, bright and dark features in these images, are observed. Scan tunneling spectroscopy reveals in gap states, implying the bright and dark defects are donor and acceptor in nature, respectively (see Supplementary Figure 6 and Supplementary Note 1 for details). Across samples, F1, F2, and F3, the total defect density increases from $2 \times 10^{11}$ to $4 \times 10^{11}$, and to $8 \times 10^{11}$ $cm^{-2}$, while the donor density *decreases* from $1.5 \times 10^{11}$ $cm^{-2}$, to $1.3 \times 10^{11}$ $cm^{-2}$, to $4 \times 10^{10}$ $cm^{-2}$. Figures 3d-e show the corresponding gate dependent photoluminescence of monolayers exfoliated from these three types of crystal. The excitonic satellites are clear in sample F1 with abundant donors but lowest total defect density, are barely visible in F2, and are not seen in sample F3. The satellites peak intensity thus track the donor density, supporting assignment of the observed photoluminescence satellites to excitonic states bound to dilute donor defects.

**Donor bound dark excitons and phonon replicas**

We examine the energy difference between the observed satellites peaks. The extracted energies of S2 and S3, relative to S1, at zero gate voltage are ($12.5 \pm 0.7$) meV and ($23.1 \pm 0.5$) meV, respectively, where the error bar is the standard deviation of the peak positions from 10 samples. These energy differences are similar to the calculated phonon energy of $K_2$ (12 meV) and $\Gamma_5$ (22 meV), respectively[26]. We thereby attribute S1 to the zero-phonon line of donor bound excitonic emission, while S2 and S3 as its $K_2$ and $\Gamma_5$ (or $E''$) phonon replicas. The assignment of S2 and S3 resembles the observed phonon replicas for free 2D excitons, pointing to the strong phonon-exciton interactions in monolayer $WSe_2$ [24-27]. The E'' phonon replica from a localized quantum emitter have also been observed in monolayer $WSe_2$,[49] albeit with different emitter properties from those reported here.

Since in $WSe_2$ the ground state configuration of a free exciton is the spin-forbidden dark exciton $X_d$,[26,27] we infer that the zero-phonon line S1 arises from a donor bound dark exciton $DX_d$. The dark exciton long lifetime makes possible efficient formation of $DX_d$, even at very low exciton density. This state should be composed by a positive charged ion, a donor bound electron, and an electron-hole pair. For the lowest energy configuration, the two electrons should have opposite spins to form spin singlet state.

**Photoluminescence excitation spectroscopy**

The above understanding of the charge configuration of $DX_d$ is further supported by polarization-resolved PL excitation spectroscopy (PLE). We tuned the energy of a continuous wave laser from 1.65 – 1.75 eV while collecting the satellite emission. As a reference, the PL spectrum obtained with 1.96 eV excitation is provided in Fig. 4a. The polarization resolved PL response of the satellites under $\sigma^+$ excitation is shown in the left ($\sigma^+$ detection) and middle ($\sigma^-$ detection) panels of Fig. 4b. The extracted PL polarization $\eta$ is shown in the right panel with the satellite PL under 1.96 eV excitation overlaid in black. Comparing Figs. 4a and 4b, we can see that satellite emission is greatly enhanced when the laser is in resonance with free neutral exciton, which produces large population of dark excitons and thus $DX_d$. The satellite emission is suppressed when resonant with the dark trion. This can be understood as the creation of dark trion competes with the formation of dark exciton, and thus donor bound dark exciton states.

A notable feature is the polarization-dependent response when the laser is in resonance with free bright trions. As shown in the $\eta$ panel in Fig. 4b, the blue regions represent PL polarization reversal when the laser is resonant with the intervalley trion ($X_T^-$), opposite to that under excitation at the intravalley resonance ($X_S^-$). This contrast is highlighted by polarization dependent spectra in Fig. 4c. The satellite emission is co- and cross-circularly polarized with laser in resonance with $X_S^-$ (E = 1.699 eV, bottom panel) and $X_T^-$ (E = 1.706 eV, top panel), respectively. Note that the vertical white stripe in $\eta$ at ~1.59 eV indicates that S3 has negligible circular polarization.

**Discussion**

**Donor spin state initialization**

Figure 5 explains the above trion excitation dependent PL polarization via donor spin state initialization process (see Supplementary Figure 7 and Supplementary Note 2 for details). Figure 5a depicts the valley-spin coupled band edges, with initially unpolarized electrons on shallow donor levels[50]. Using resonant excitation of intervalley trion $X_T^-$ in Fig. 5b as an example, $\sigma^+$ polarized excitation creates an electron and hole pair in the $K$ valley, which pairs with a spin up electron in the $K'$ valley and forms $X_T^-$. This optical pumping process depletes the spin up electrons associated with donors in $K'$, eventually resulting in a net population of spin down electrons in the K valley bound to the shallow donors (denoted as $D_{K,\downarrow}$). $D_{K,\downarrow}$ then selectively binds to a dark exciton in the $K'$ valley ($X_d^{K'}$), because of the Pauli exclusion. The donor bound state emission can happen via either defect assisted direct electron-hole recombination yielding S1 PL peak, or phonon assisted stokes emission (either $K_2$ for S2 peak or $\Gamma_5$ for S3 peak), via an intermediate bright trion state (Fig. 5c). In all cases, the emission is $\sigma^-$ polarized as determined by the hole valley configuration, cross-polarized to the excitation[26].

When $\sigma^+$ polarized excitation is in resonance with intravalley trion $X_S^-$, it depletes instead the spin down electrons in $K$ valley, and leaves a net population of spin-up electron in $K'$ valley bound to the shallow donor ($D_{K',\uparrow}$), as shown in Fig. 5d. This leads to donor bound dark exciton states ($DX_d^K$) with spin-valley polarization opposite to that in resonant excitation of $X_T^-$. The donor bound state then emits $\sigma^+$ polarized light, co-polarized with the excitation (Fig. 5e). We note that, although pump is $\sigma^+$ polarized, supply of free neutral dark exciton is expected in both valleys because of the ultrafast valley depolarization through electron-hole exchange in the formation of neutral excitons, so the $DX_d^K$ spin-valley polarization is determined by the optical orientation of donor electron. We would also like to point it out that although S2 behaves similarly to S1, which both have strong circular polarization, $S_3$ is unpolarized. This suggests S3 maybe not be as simple as a $\Gamma_5$ phonon replica of S1, which needs further investigation.

## Origin of intrinsic defect

Our results demonstrated the observation of excitons bound to intrinsic defects. Extrinsic defects (e.g. O at Se site) and confinement potential, which can be responsible for localized single photon emitters with strong optical anisotropy[32-38], are thus not candidates. There are four types of intrinsic defects: W vacancy, Se vacancy, W replace Se site, and Se replace W site. Although Se vacancy has been suggested to be responsible for observed broad low energy PL features[51], our atomically resolution STM measurements (Supplementary Figure 6e) rules out vacancy as donor candidates. In addition, previous study on similar crystals with low defect density found the chalcogenide vacancy to be very rare[39]. Calculation also suggested Se at W antisite ($Se_W$) is a deep defect with multiple in gap states[32], and the exciton bound to $Se_W$ is expected to be about 300 meV below the $WSe_2$ A exciton.[32] All these are distinct from our experimental observation: our STS on donor (Supplementary Figure 6a) only shows one shallow defect band near the CBM, while the donor bound exciton is about 120 meV below the A exciton. The leaves W at Se antisite ($W_{se}$) as a possible candidate. However, calculation suggests that $W_{se}$ is an acceptor, which has multiple charge levels with a band near CBM[52]. This is distinct from our results that the defect is donor type with only one charge level within the gap. The current study cannot resolve the exact type of the defect, which requires further experimental and theoretical efforts (see discussion).

## Outlook

Our results reveal light emission from dark exciton bound to intrinsic dilute donors and possible phonon replicas in the ultraclean monolayer $WSe_2$. Similar behavior is observed in III-V semiconductors[53], where defect density must be low before individual defect-bound exciton peaks and charging states are resolvable. It remains to be seen whether the defect bound excitons exist in multilayer samples and other transition metal dichalcogenides. The observed long population and polarization lifetimes are advantageous for exploiting the spin-valley functionalities using monolayer $WSe_2$. The possibility of probing single-donor emitting sites by further reducing the native defect density, or by using near-field techniques, is an interesting direction towards realizing optical spin quantum memory in 2D materials. The PLE results are promising in this respect, since they demonstrate optical initialization of the donor-bound electron spin state by selective excitation of different trion states. On the other hand, interactions between neighboring donor sites may lead to interesting many-body behaviors, such as long-range correlation. Non-classical photon statistics are expected to emerge in both regimes, although weak oscillator strength may pose significant challenges. Our work prompts research efforts to identify and engineer the underlying defect and its electronic configuration, and we expect that further improvements in sample quality will enable detailed studies of the satellite fine structure, such as vibrational and rotational spectrum, hyperfine interactions, excited states, and optical orientation by magnetic fields control of donor spin states.

**Methods:**

**Crystal Growth.** For growth of $WSe_2$ crystals with varying defect densities (Fig. 3), $WSe_2$ crystals were synthesized by reacting W with Se flux. W powder (99.999%) and Se shot (99.999%) mixed in 1:100 (F1), 1:15 (F2), and 1:5 (F3) atomic ratios, were loaded into quartz ampules separately which were then evacuated and sealed at ~$10^{-6}$ Torr. The ampules were vertically placed into a box furnace and heated to 1080 °C over 48 hours. After a dwelling time of 1 week at 1080 °C, it was slowly cooled down to 300 °C at a rate of 0.6 °C/h. The obtained $WSe_2$ crystals were subsequently filtered from the Se flux by quartz wool and annealed at 275 °C for 24 hours in a vacuum quartz ampule.

**Sample Fabrication**. The van der Waals heterostructure samples used in this study were fabricated by polycarbonate-based viscoelastic dry-transfer techniques. The polymer residues were cleaned by baths in chloroform and isopropyl alcohol. The individual layers were obtained from the mechanical exfoliation of bulk crystals onto 285 nm of thermally-grown $SiO_2$ on p+ doped Si wafers. The thickness of hBN layers was determined by atomic force microscopy. Monolayers of $WSe_2$ were identified by their optical contrast, which was later confirmed by their low energy PL spectrum.

**Electrostatic Doping**. Standard electron beam lithography was used to produce PMMA masks for subsequent electron beam evaporation of V/Au (nominally, 5/50 nm) electrodes to the graphite backgates, as well as to a small portion of the monolayer $WSe_2$ protruding from underneath the top gate dielectric. For fully-encapsulated monolayer $WSe_2$ samples, electrodes were deposited to a second piece of thin graphite that was both in contact with the monolayer $WSe_2$ and also protruding from underneath the top gate dielectric. The applied backgate voltage was controlled by analog output from a National Instruments USB I/O DAQ board using mxdaq drivers in Matlab environment. The gate leakage current was actively monitored using a combination of transimpedance amplifier and analog input on the DAQ board.

**Photoluminescence Spectroscopy.** Photoluminescence measurements were performed in a home-built confocal microscope, in reflection geometry, normal to the plane defined by the monolayer $WSe_2$. The samples were either (1) mounted on the cold head of a closed-cycle He cryostat at a temperature of 5 K and studied using an IR-enhanced achromatic 50X objective lens (0.65 NA) or (2) mounted inside a He-exchange-gas cooled cryostat (attocube attoDRY 2100) at a temperature of 1.6 K and studied using an IR-enhanced achromatic 100X objective (0.81 NA) with non-magnetic Ti housing. In both cases, the samples were illuminated by power-stabilized HeNe laser light ($\lambda$ = 1.96 eV) focused to a beam waist of approximately 1 μm. Circularly-polarized excitation and detection was achieved by an appropriate combination of fixed linear polarizers and $\lambda/4$-waveplates, with achromatic $\lambda/2$-waveplates mounted in stepper-motor-controlled rotation stages. Linearly-polarized excitation and detection was achieved by an appropriate combination of fixed linear polarizers and $\lambda/2$-waveplates mounted in stepper-motor controlled rotation stages. The collected PL was directed into a 0.5 m spectrometer, where it was dispersed by a 600 line/mm grating with 750 nm blaze before being detected by Si charge-coupled-device. The polarization of the light entering the spectrometer was S-polarized for all polarization-resolved measurements.

**Photoluminescence Excitation Spectroscopy.** The excitation source was a narrowband (< 20 kHz linewidth) and frequency tunable Ti:Sapphire continuous-wave laser ($M^2$ SolsTiS). The laser line was filtered from the collection path using a combination of bandpass filter and tunable long- and short-pass filters (Semrock VersaChrome). The optical power was stabilized by servo control of the power in the first order diffracted beam using feedback control of the voltage applied to the acoustic-optic modulator.

**Time Resolved Photoluminescence Spectroscopy.** Time-resolved PL measurements were performed by directing the collected photons onto an IR-enhanced single-photon avalanche photodiode (Excelitas) connected to a time-correlated single photon counting system (PicoHarp 400). Spectral filtering of the signal was achieved by either (1) a combination of bandpass filter and tunable long- and short-pass filters (Semrock VersaChrome), or (2) using the 0.5 m monochromator to disperse the signal, which was then filtered at the exit port by an adjustable slit assembly. The excitation was provided by spectrally filtered (~2 nm bandwidth at FWHM) output from a supercontinuum fiber laser with 100 kHz repetition rate and approximately 10 ps pulse duration.

**Scanning Tunneling Microscopy (STM).** STM measurements were performed using a Scienta Omicron STM system at room temperature under an ultra-high vacuum (base pressure < 1.0 x 10-10 torr). $WSe_2$ bulk crystals were mounted onto a metallic sample holder with silver epoxy, and then cleaved *in situ* in the UHV STM chamber to obtain a clean surface. The tungsten tip was cleaned and calibrated against an Au(111) surface before all the measurements. For each STM image, the defect density was calculated through the number of defects divided by the scanning size. To avoid the localized counting, each value was obtained from the average of 30 (50x50 $nm^2$) STM images in different area.

**Acknowledgments:** This work was mainly supported by Army Research Office (ARO) Multidisciplinary University Research Initiative (MURI) program (grant no. W911NF-18-1-0431) and NSF EFRI (Grant No. 1741656). Time resolved measurements are supported by the Department of Energy, Basic Energy Sciences, Materials Sciences and Engineering Division (DE-SC0018171). HM acknowledges support from a Samsung Scholarship. DE acknowledges partial support from the NSF Center for Integrated Quantum Materials (CIQM). WY and HY were supported by the Research Grants Council of Hong Kong (C7036-17W), and the University of Hong Kong Seed Funding for Strategic Interdisciplinary Research. DM and JY were supported by the US Department of Energy, Office of Science, Basic Energy Sciences, Materials Sciences and Engineering Division. $WSe_2$ synthesis and STM characterization were supported by the NSF

MRSEC program through Columbia in the Center for Precision Assembly of Superstratic and Superatomic Solids (DMR-1420634). KW and TT acknowledge the support from the Elemental Strategy Initiative conducted by the MEXT, Japan, Grant Number JPMXP0112101001, JSPS KAKENHI Grant Numbers JP20H00354 and the CREST(JPMJCR15F3), JST. HD is supported by the Department of Energy, Basic Energy Sciences, under Award No. DE-SC0014349. XX acknowledges the support from the State of Washington funded Clean Energy Institute and from the Boeing Distinguished Professorship in Physics.

**Author Contributions:** XX, WY, JH, DE, AP supervised the projects. PR and MH fabricated the devices, performed optical spectroscopy measurements, and firstly identified the excitonic satellites with bulk crystal provided JY and DGM. BK, SL, and DAR performed defect control $WSe_2$ crystal synthesis, device fabrication, STM and optical spectroscopy measurements and analysis. CRV performed and analyzed atomically resolution STM/STS measurements. HM and LM performed super resolution measurements. KW and TT provided high quality hBN crystal. XX, WY, PR, HY, JH and HD interpreted the results with input from all authors. XX, PR, MH, and WY wrote the paper with inputs from all authors. All authors discussed the results.

**Competing Interests:** The authors declare no competing interests.

**Data Availability**: The data that support the findings of this study are available from the corresponding authors upon reasonable request.

## Figures

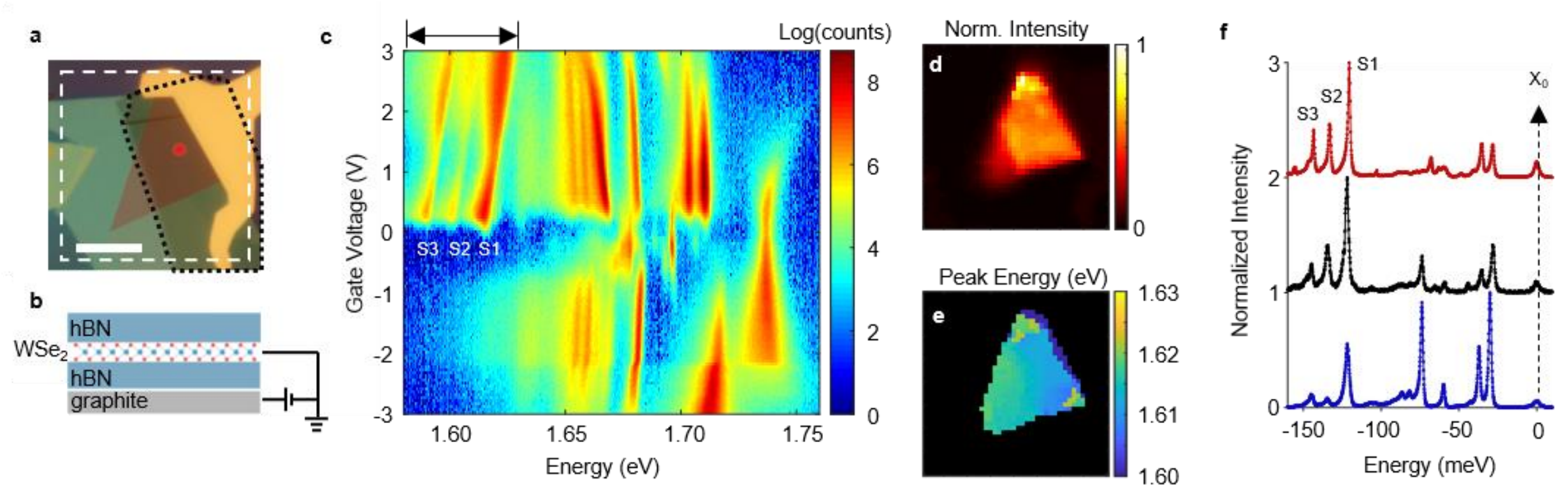

**Figure 1| Emergence of deeply bound exciton satellites | a,** Optical image of electrostatic gating device composed of exfoliated monolayer $WSe_2$ (red area) encapsulated in hBN with graphite backgate (outlined in black). Scale bar is 10 µm. **b,** Schematic of sample side view. **c,** Photoluminescence as a function of back gate voltage with the laser spot indicated by the red dot in **a**. Three satellites peaks near 1.60 eV appear when the sample is n-doped. **d,** Spatial map of the integrated PL from the satellite peaks at $V_b$= 0.5 V. Integration spectral region shown by arrows on top of **c**, and spatial region outlined by dashed white lines in **a**. **e,** Spatial map of the peak energy of the highest energy satellite at $V_b$= 0.5 V. **f,** Waterfall plot of PL spectra from 3 different samples, showing homogeneous satellite binding energies and robust three peak spectral features. The energy axis is scaled relative to the free neutral exciton ($X_0$ at ~1.735 eV).

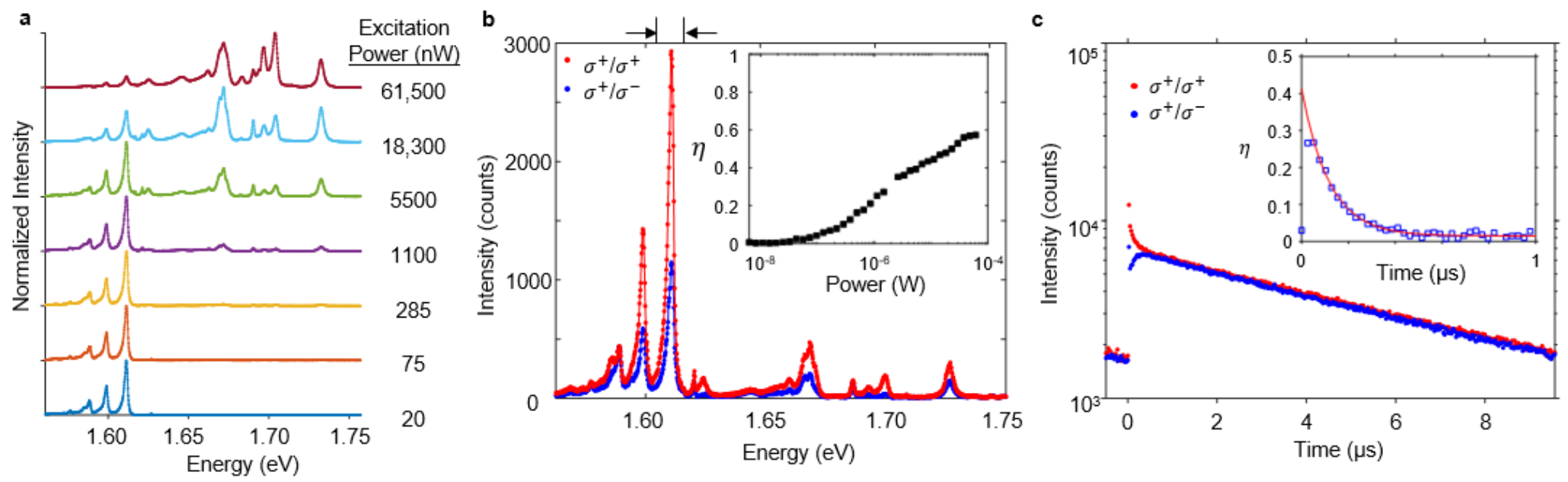

**Figure 2 | Power, polarization, and time resolved PL from satellites. a,** Waterfall plot of normalized PL spectra at selected excitation powers. The exciton satellites dominate the spectrum at low powers. **b,** Polarization resolved spectrum of satellite peaks under $\sigma^+$circularly polarized excitation (1 µW). Inset: Power dependence of $\eta$, the degree of circular polarization, which grows with increasing power up to $\eta \approx 0.6$. **c,** Time resolved PL of S1 reveals a population lifetime of 6.95 $\pm$ 0.05 µs and a polarization lifetime of 116 $\pm$ 4 ns (red line is single exponential fit). All data are from Device 2 at $V_b$ = 0 V. Top arrows in **b** indicate the spectral region of integration.

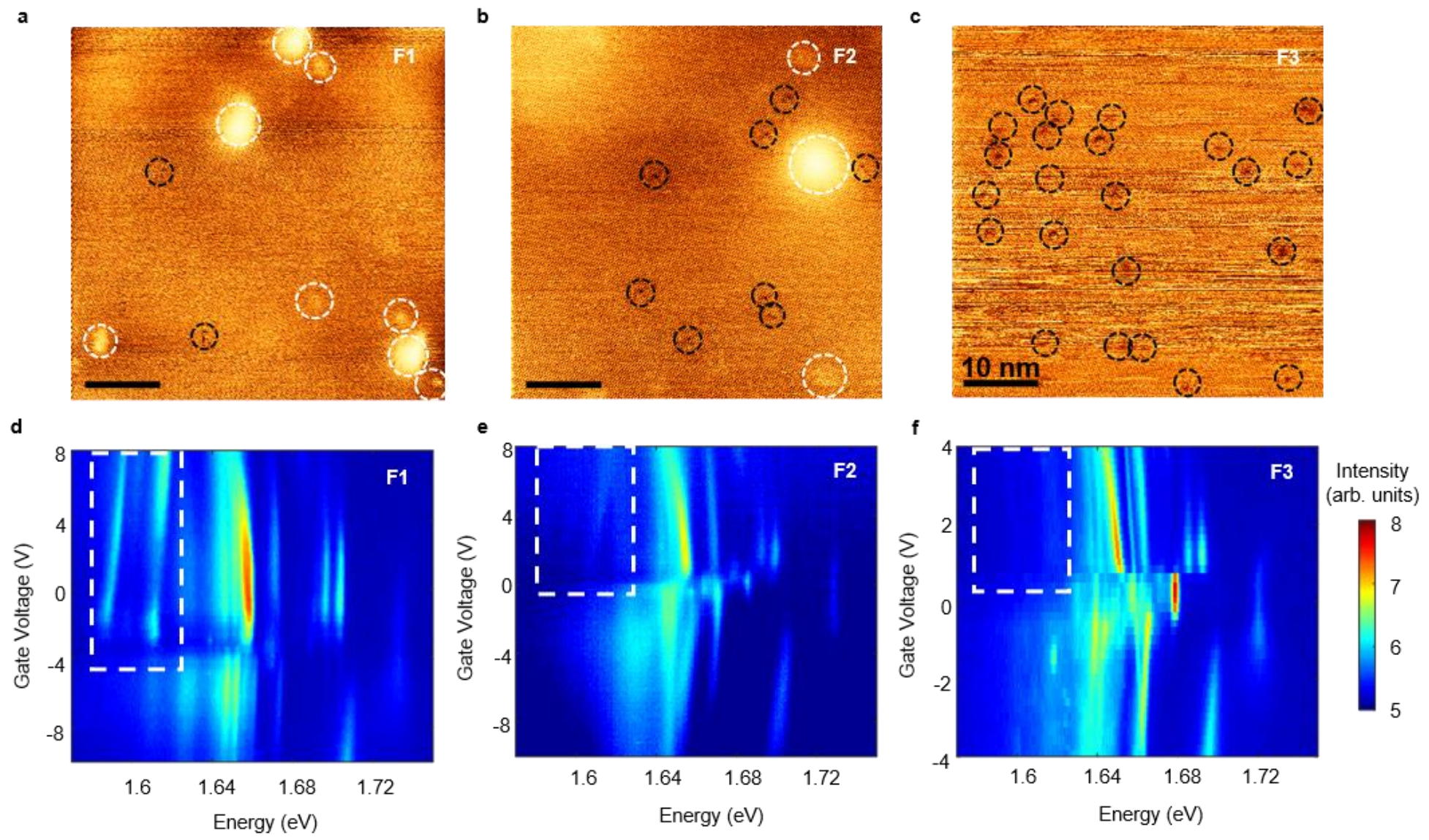

**Figure 3 | Characterization of defects and defect bound excitons in $WSe_2$. a-c,** Scanning Tunneling Microscopy (STM) topographic images of 50x50 $nm^2$ area of $WSe_2$ with different growth parameters (F1, F2, and F3). Scale bar is 10nm. Imaging conditions for the STM topographic images were a tunneling bias of 1.4 V and current of 400 pA for F1 and a tunneling bias of 1.4 V and current of 150 pA for F2 and F3. The $WSe_2$ crystals were cleaved in ultra-high vacuum STM chamber (base pressure < $2.0\times10^{-10}$ torr) to obtain a clean surface before imaging. White and black dashed circles highlight the donor and acceptor defects, respectively. See Supplementary Figure 6 details. **d-f,** Photoluminescence (PL) color maps as a function of the back-gate bias for **d,** F1, **e,** F2, and **f,** F3. The white dashed boxes in the PL color maps indicate the energy and the back-gate bias ranges for defect bound excitons. For the PL color map, the samples were excited using a continuous-wave (CW) laser with an excitation wavelength of 532 nm with a fluence of 650 W/$cm^2$ (10 μW) at 4 K.

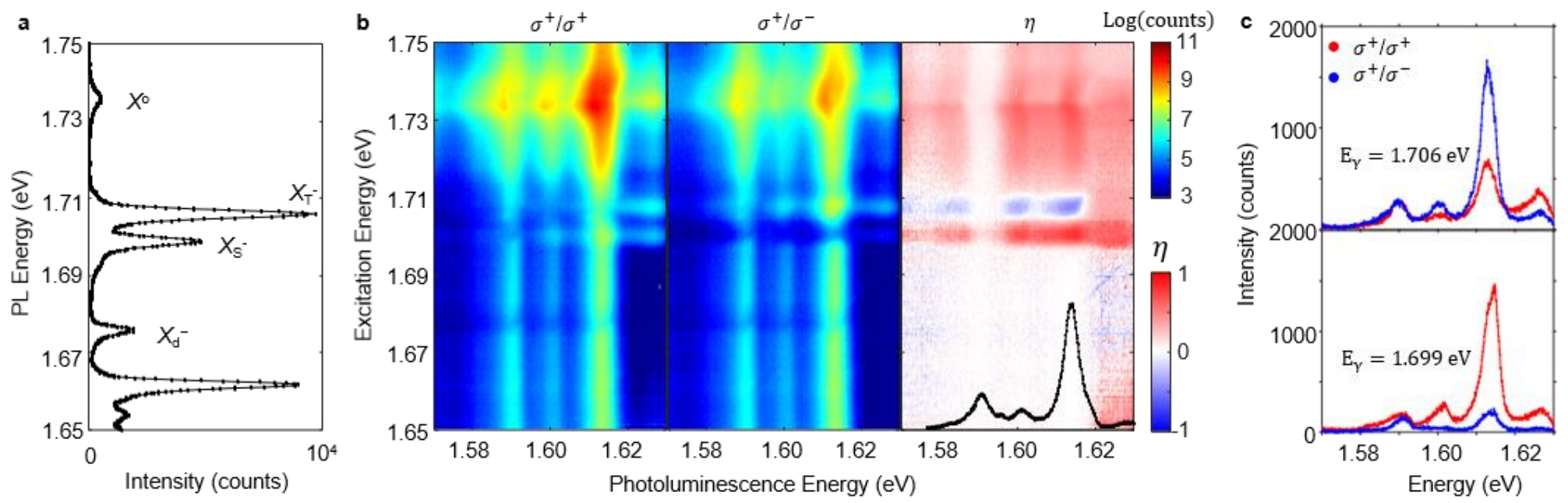

**Figure 4 | Excitation energy dependent spin-valley depletion and PL polarization reversal.** All data are from Device #1 at $V_b$ = 0.5 V. **a,** PL spectra with a HeNe laser excitation. **b,** Polarization resolved PL spectra by sweeping $\sigma^+$polarized excitation from 1.75 to 1.65 eV. The $\sigma^+$ and $\sigma^-$ components of the PL are shown in the left and middle panels, respectively, while the right panel shows the extracted degree of polarization $\eta$. **c,** Polarization dependent PL with the excitation in resonance with the free intervalley (top, 1.706 eV) and intravalley (bottom, 1.699 eV) trions, showing polarization reversal of the exciton satellite PL.

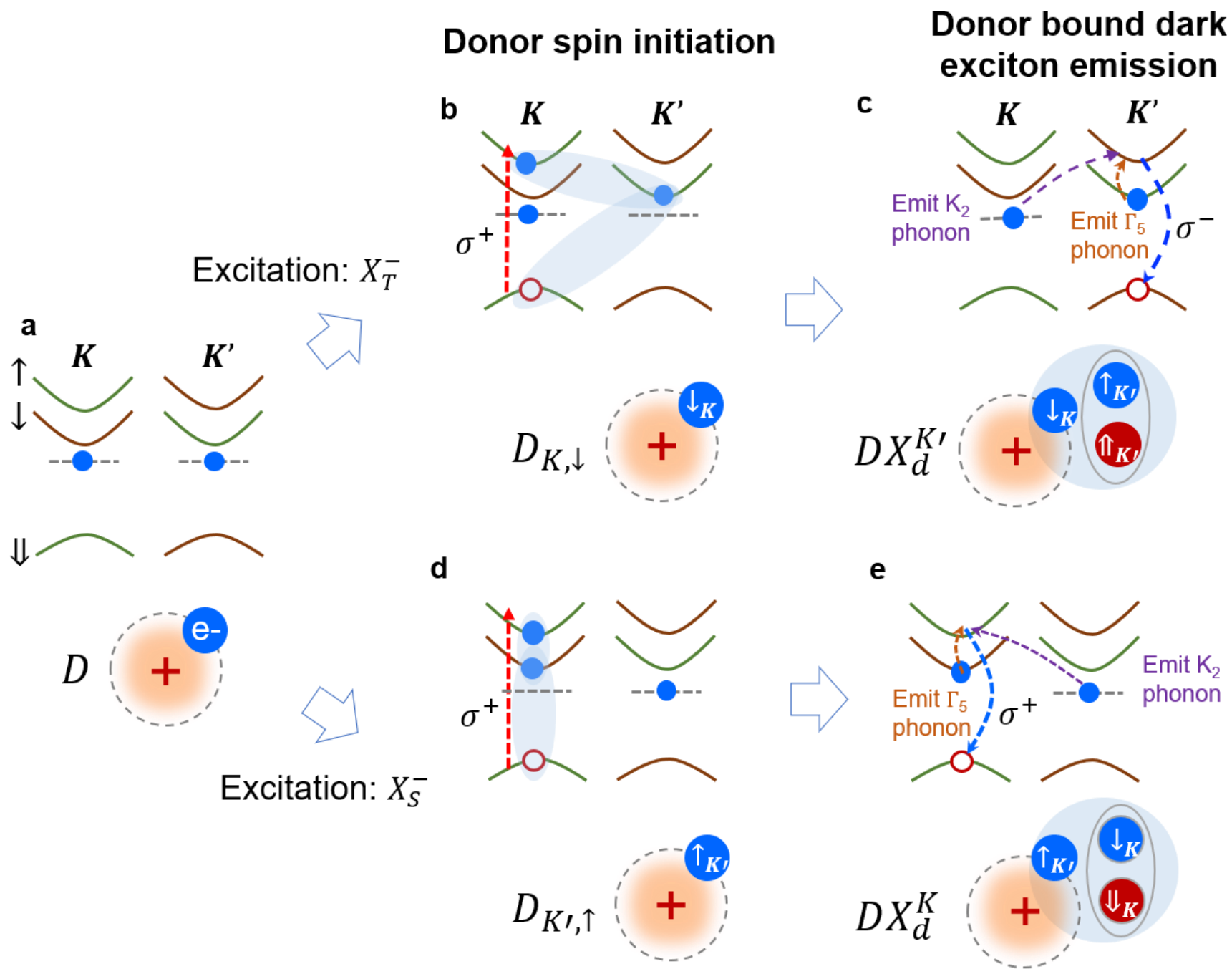

**Figure 5 | Schematic of donor bound dark excitons and spin state initialization. a,** Top is a schematic of the spin-valley coupled band edges. Green and brown lines denote electron (hole) spin pointing up (down) and down (up), respectively. Dashed lines indicate a shallow donor level, where carrier's spin-valley locked index is preserved in the relatively smooth trapping potential. Bottom cartoon illustrates the donor configuration $D$ where an electron (blue circle) is trapped by a positive charge center. Under σ+ polarized resonant excitation of free intervalley trion $X_T^-$, **b,** the formation of $X_T^-$ (top cartoon) depletes the donor electron spins in $K'$ valley. This results in optical orientation of spin and valley polarized electron coupled to the shallow donor (bottom cartoon, $D_{K,\downarrow}$). $D_{K,\downarrow}$ can then capture a neutral dark exciton in the opposite valley, **c,** forming a donor bound dark exciton $DX_d^{K'}$. $DX_d^{K'}$ can emit light via defect assisted direct electron-hole recombination (peak S1), or via coupling to the bright trion assisted by phonons, which are by either emitting valley conserved but electron spin flipped $\Gamma_5$ phonon (peak S3), or emitting electron spin conserved but valley flipped $K_2$ phonon (peak S2). The opposite hole valley configuration of $X_T^-$ and $DX_d^{K'}$ dictates that the emitted light of $DX_d^{K'}$ is σ- polarized and opposite to that of excitation. **d-e**, illustrates the scenario of resonant excitation of free intravalley $X_S^-$, which depletes spin down electrons of donor in the same valley. The donor bound dark exciton formed then has the same hole valley configuration, and emit light of the same circular polarization as the excitation. See text for details. Note that the schematics used here are based on single particle picture, for the convenience of explaining spin, valley, and charge degrees of freedom of the quasiparticles, and the stokes process with different phonons.

# Supplementary Information for
# Title: Intrinsic Donor Bound Excitons in Ultraclean Monolayer Semiconductors

**Authors:** Pasqual Rivera[1], Minhao He[1], Bumho Kim[2], Song Liu[2], Carmen Rubio-Verdú[3], Hyowon Moon[4], Lukas Mennel[4], Daniel A. Rhodes[2], Hongyi Yu[5], Takashi Taniguchi[6], Kenji Watanabe[7], Jiaqiang Yan[8,9], David G. Mandrus[8-10], Hanan Dery[11], Abhay Pasupathy[3], Dirk Englund[4], James Hone[2#], Wang Yao[5#], & Xiaodong Xu[1,12#]

**Affiliations:**

[1]Department of Physics, University of Washington, Seattle, Washington 98195, USA

[2]Department of Mechanical Engineering, Columbia University, New York, NY, 10027, USA

[3]Department of Physics, Columbia University, New York, NY, 10027, USA

[4]Department of Electrical Engineering and Computer Science, Massachusetts Institute of Technology, Cambridge, MA, 02139, USA

[5]Department of Physics, University of Hong Kong, and HKU-UCAS Joint Institute of Theoretical and Computational Physics at Hong Kong, China

[6]International Center for Materials Nanoarchitectonics, National Institute for Materials Science, Tsukuba, Ibaraki 305-0044, Japan.

[7]Research Center for Functional Materials, National Institute for Materials Science, Tsukuba, Ibaraki 305-0044, Japan.

[8]Materials Science and Technology Division, Oak Ridge National Laboratory, Oak Ridge, Tennessee, 37831, USA

[9]Department of Materials Science and Engineering, University of Tennessee, Knoxville, Tennessee, 37996, USA

[10]Department of Physics and Astronomy, University of Tennessee, Knoxville, Tennessee, 37996, USA

[11]Department of Electrical and Computer Engineering, University of Rochester, Rochester, New York, 14627, USA

[12]Department of Materials Science and Engineering, University of Washington, Seattle, Washington, 98195, USA

[#]Correspondence to xuxd@uw.edu; wangyao@hku.hk; jh2228@columbia.edu

**Supplementary Figures:**

1. **Assignment of free 2D exciton states**
2. **Spatial homogeneity of the satellite emission**
3. **Robust satellite properties across many samples**
4. **Power dependence from additional samples**
5. **Linear polarization resolved satellite emission**
6. **STM/STS measurements of $WSe_2$ crystals**
7. **Trion excitation dependent donor bound exciton photoluminescence**

**Supplementary Notes:**

1. **Assignment of donor type defect**
2. **Trion excitation dependent donor bound exciton photoluminescence**

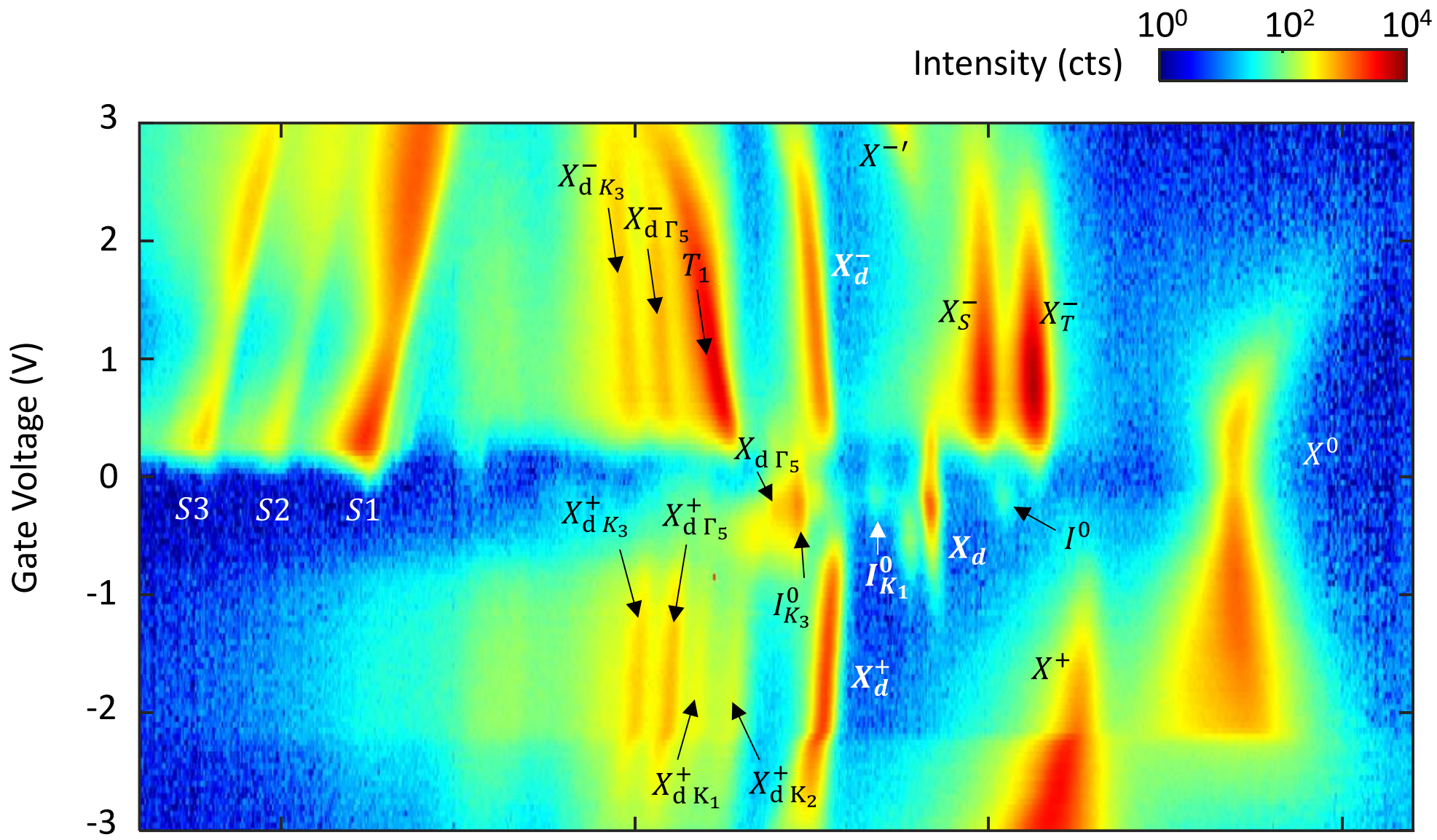

**Supplementary Figure 1. Assignment of free 2D excitons in monolayer WSe2.** This is the same data as in Fig. 1c in the maintext, with all reported excitonic feature identified and labeled. At hole doping side (negative gate voltage), from high energy to low energy, the identified excitonic features are: positively charged trion ($X^+$),[1] positively charged dark trion ($X_d^+$),[2,3] and $K_2$, $K_1$, $\Gamma_5$, and $K_3$ phonon replica of $X_d^+$ ($X_{d\,K2}^+$, $X_{d\,K1}^+$, $X_{d\,\Gamma5}^+$, $X_{d\,K3}^+$).[4-8] At charge neutral regime, from high energy to low energy, the identified excitonic feature are: bright exciton ($X^0$),[1] intervalley dark exciton ($I^0$),[6,7] dark exciton ($X_d$),[9-14] $K_1$ and $K_3$ phonon replica of $I^0$ ($I_{K_1}^0$, $I_{K_3}^0$),[6] and $\Gamma_5$ phonon replica of dark exciton $X_{d\,\Gamma_5}$.[4-8] At electron doping side (positive gate voltage), from high energy to low energy, the identified excitonic features are: negatively charged intervalley ($X_T^-$) and intravalley trion ($X_S^-$),[15,16] $X^-{}'$ state,[1] dark trion ($X_d^-$),[2,3,9-14] $T_1$ state, [4-14] $\Gamma_5$ and $K_3$ phonon replica of $X_d^-$ ($X_{d\,\Gamma5}^-$, $X_{d\,K3}^-$).[4-8] Note that the nature of the $X^-{}'$ state and the $T_1$ state are still under active investigation.

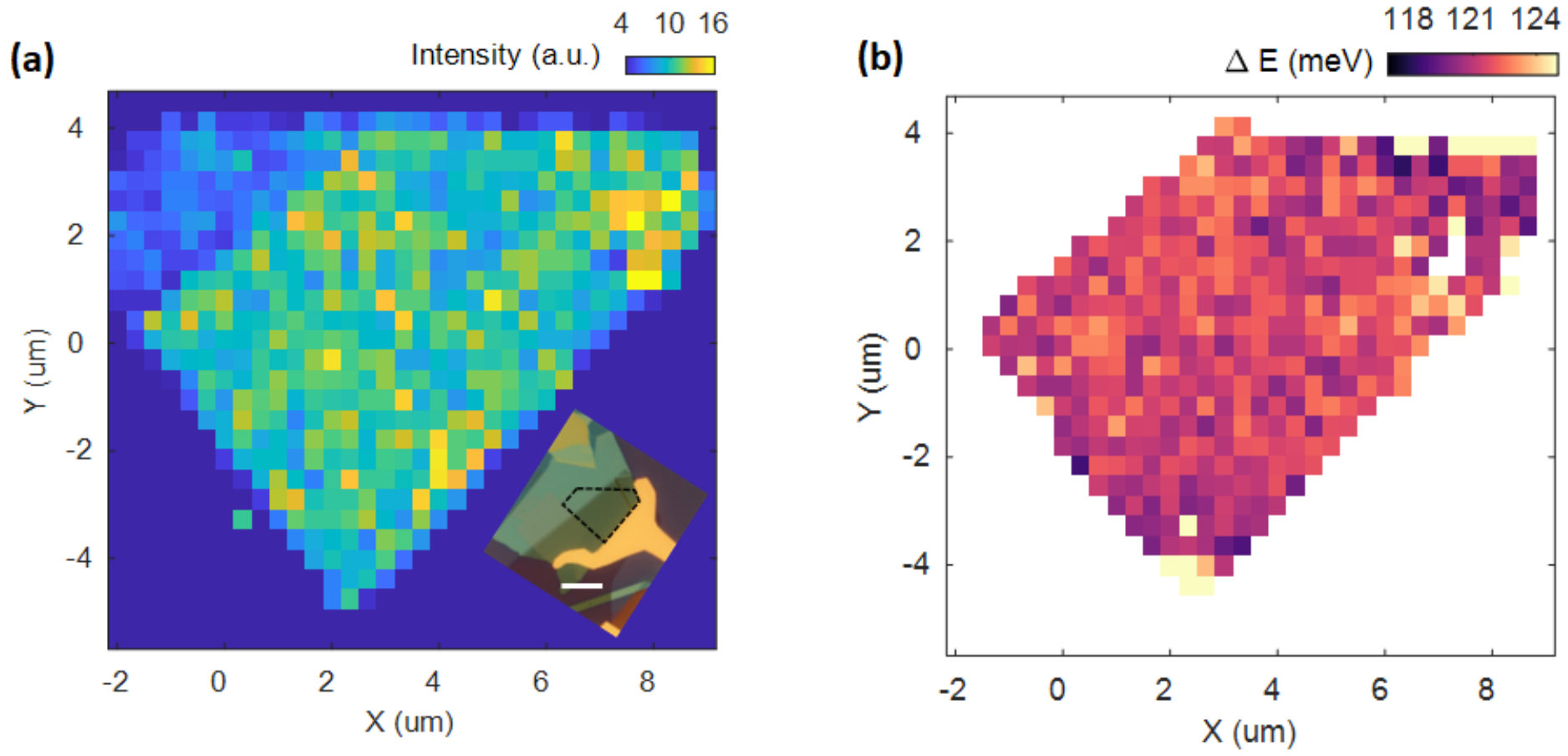

**Supplementary Figure 2. Spatial homogeneity of excitonic satellite emission. a**, Hyperspectral spatial map of satellite emission intensity, with spatial resolution of 333nm. Inset shows the optical image of the device with same orientation. Monolayer $WSe_2$ region is marked out by the black dashed line. Scale bar is 10μm. **b**, Hyperspectral spatial map of the energy difference between the first satellite peak (S1) and $X^0$.

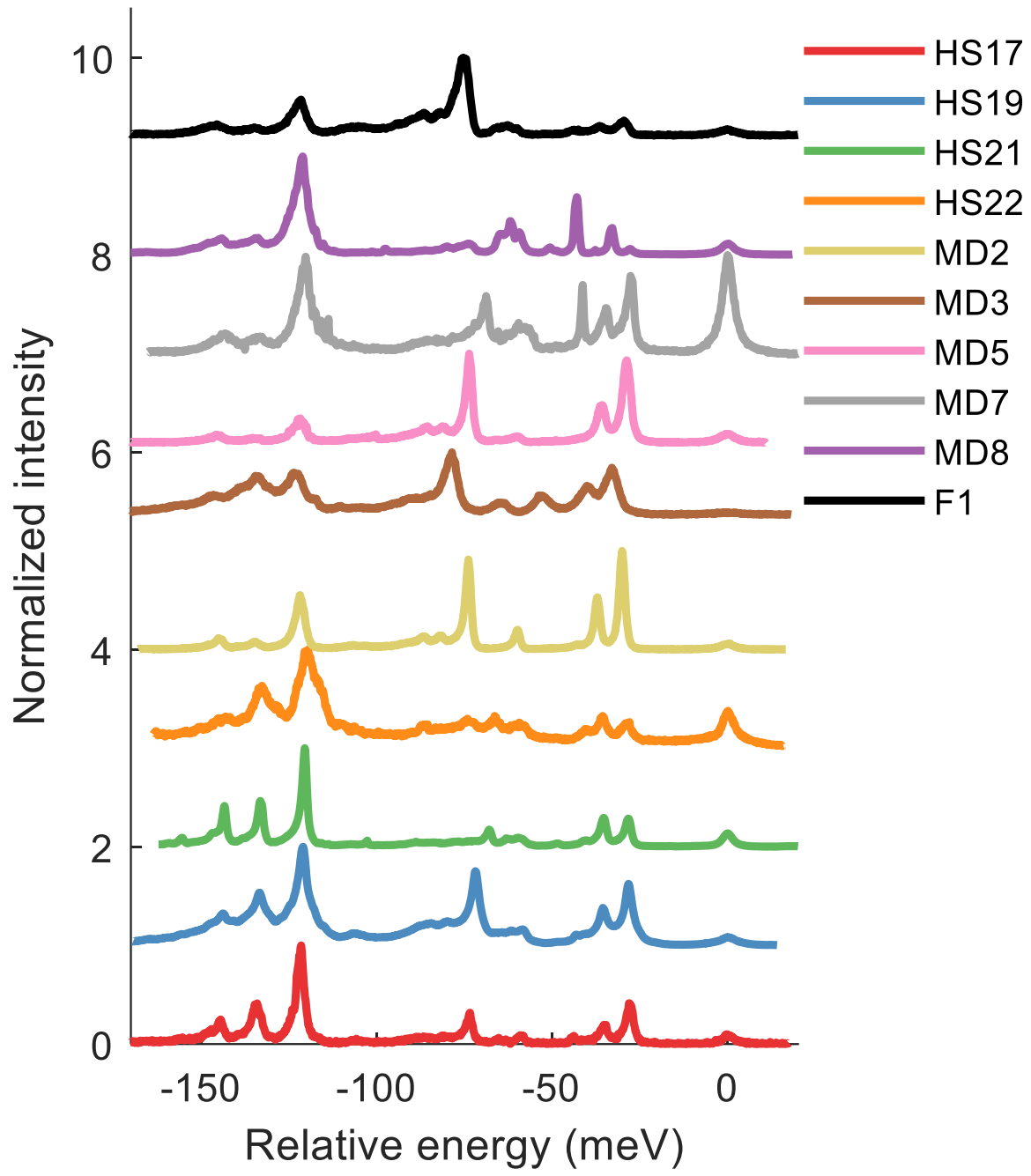

**Supplementary Figure 3. Robust satellite properties across 10 samples.** Waterfall plot of photoluminescence spectra from 10 different samples, showing homogeneous satellite binding energies and spectra structures. The energy axis is scaled relative to the neutral exciton. Spectrum are taken with no electrostatic gating and all flakes are slightly electron doped due to dilute donors. All samples were fabricated over a 3-year span. Devices HS17, HS19, HS21, HS22, MD2, MD5, MD7 were fabricated from flakes that are exfoliated from multiple crystals grown by vapor transport at ORNL. Devices F1, MD8, and MD3 were fabricated from flakes that are exfoliated from three different batches of crystals by flux growth at Columbia University.

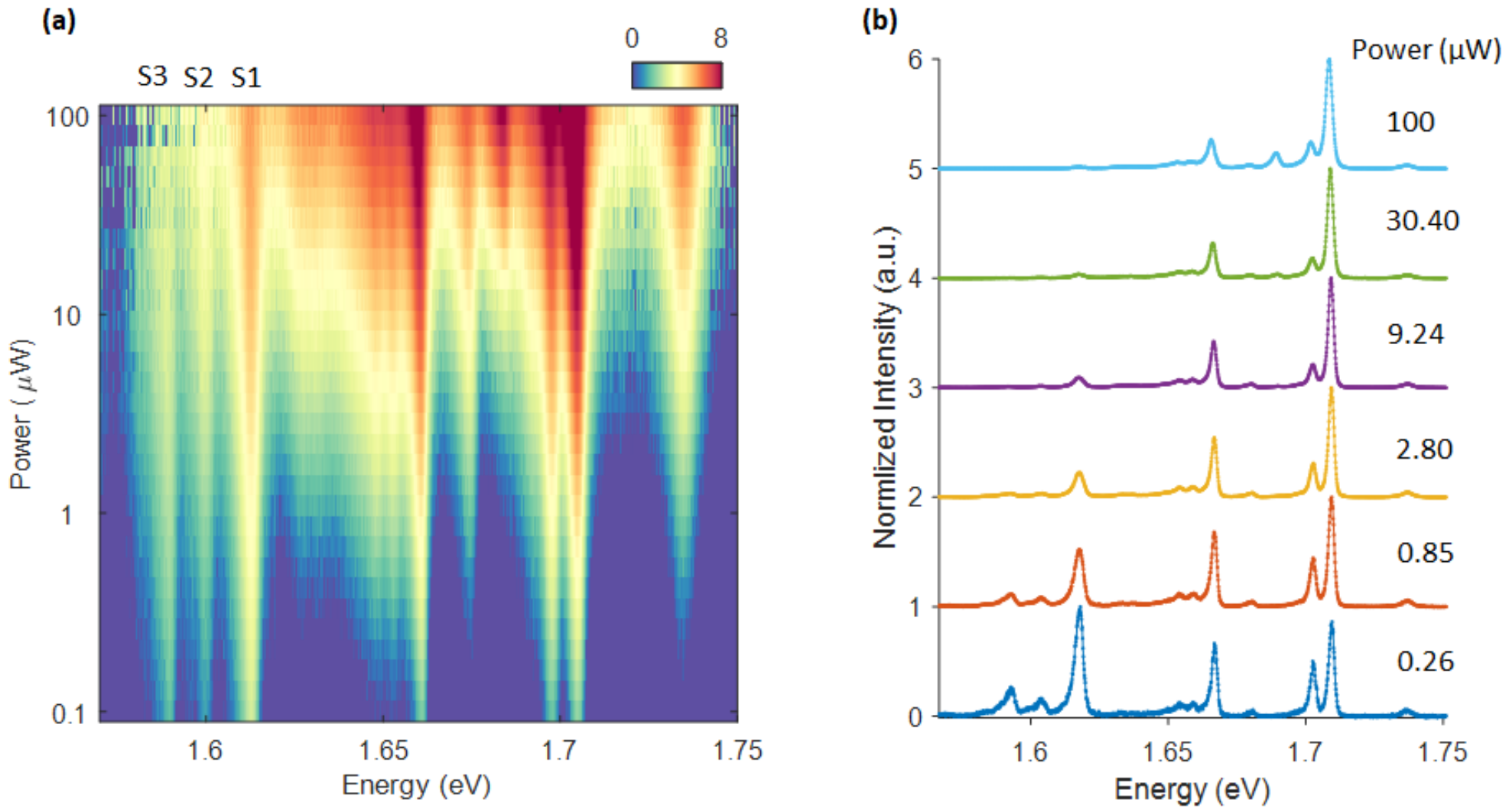

**Supplementary Figure 4. Power dependence from additional samples. a,** Photoluminescence spectrum as function of excitation power (log scale) in Device 1, at gate voltage of 0.8V. **b,** Waterfall plot of normalized PL spectra at selected excitation powers in Device 1.

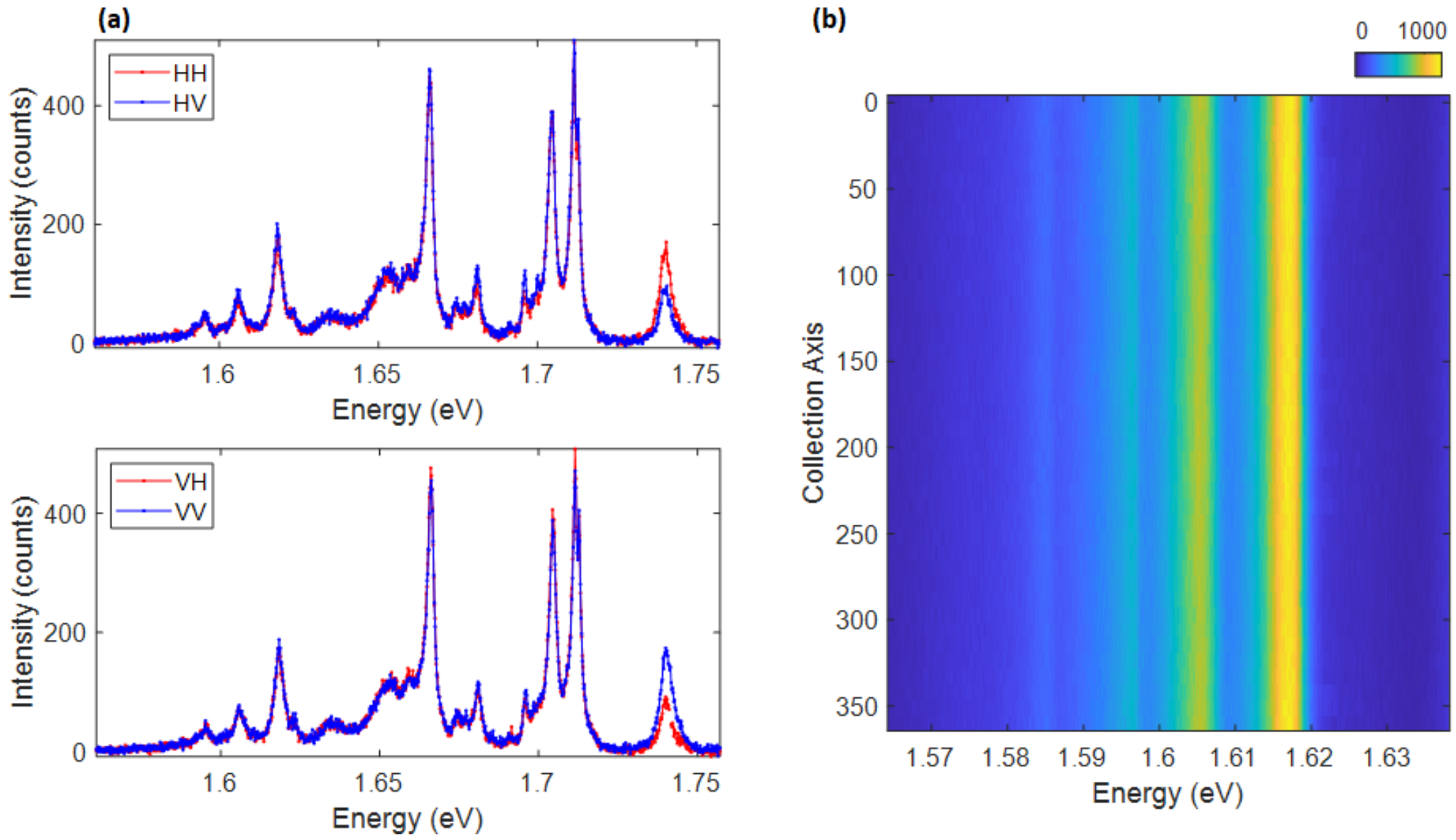

**Supplementary Figure 5. Linear polarization resolved satellite emission. a,** Photoluminescence spectra with linearly polarized excitation and collection. Data were taken from device HS 17 with 10 μW 633nm CW laser excitation. Top and bottom panels are horizontally (H) and vertically (V) polarized excitation, respectively. **b,** Satellite photoluminescence intensity plot under horizontally polarized excitation with rotating linear polarization collection. Data were taken from device HS19 with 1.5 μW 724nm CW laser

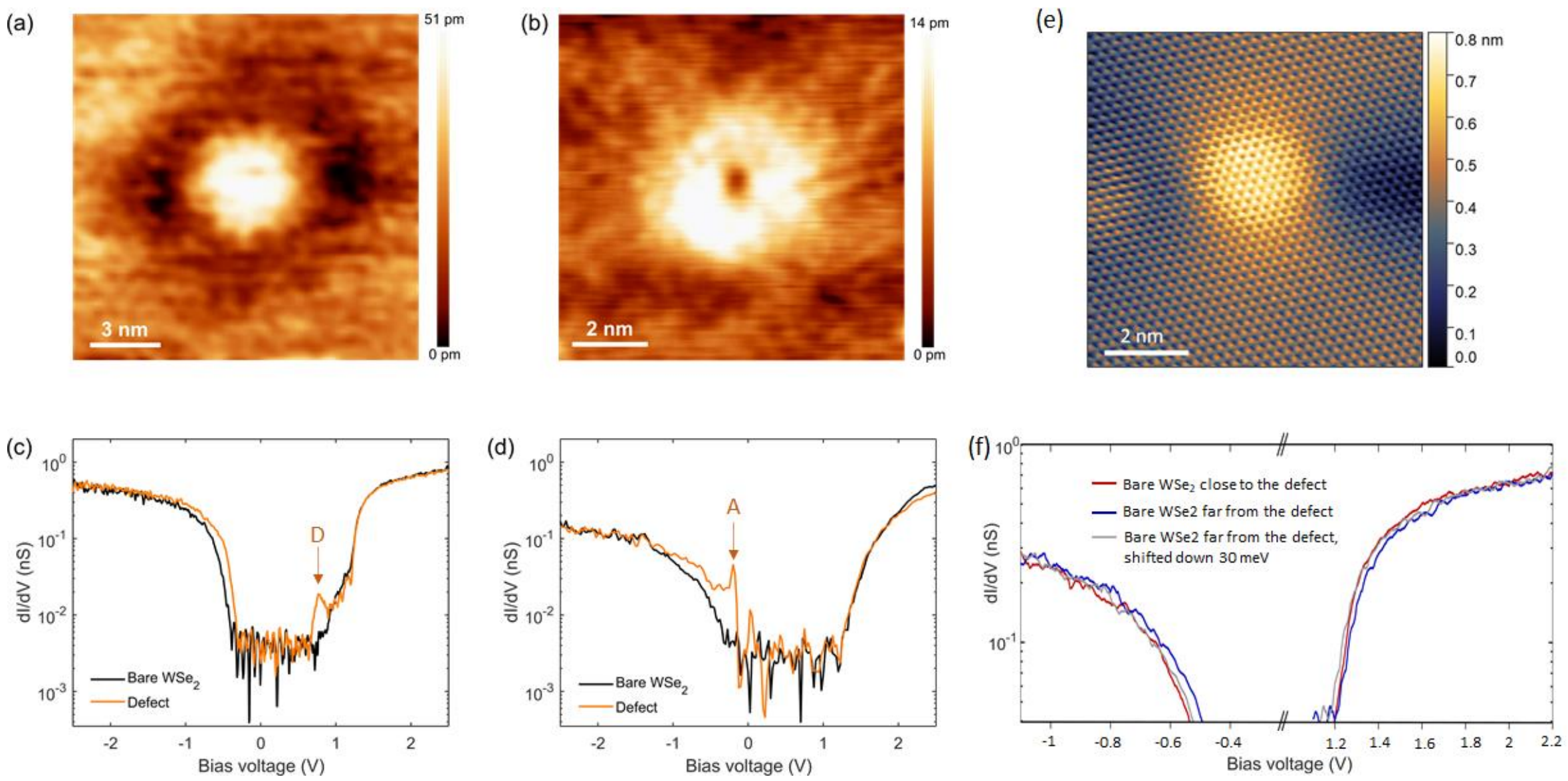

**Supplementary Figure 6. Scanning tunneling microscopy and spectroscopy analysis of donor and acceptor defects of $WSe_2$ bulk crystals at ~10K.** STM topographic images of **a**, a donor defect (imaging conditions: 1.4 V, 50 pA) and **b**, an acceptor defect (imaging conditions: 2.0 V, 50 pA). Here, (a) and (b) correspond to the bright and dark defects shown in Fig. 3 in the main text. Differential conductance curves obtained on **c**, a donor defect and **d**, an acceptor defect. Here, D and A denote donor and acceptor bands, respectively. The differential conductance obtained on the bare $WSe_2$ (black curve) is shown for comparison. The binding energies corresponding to donor and acceptor defects are about 110 meV and 125 meV, respectively. **e**, Atomic resolution STM image of $WSe_2$ bright donor defect (V=1.4V, I=100 pA) at room temperature. **f**, STS obtained on the $WSe_2$ surface close (red) and far (blue) from the donor defect. A 30 meV shifted STS is shown in grey line for comparison.

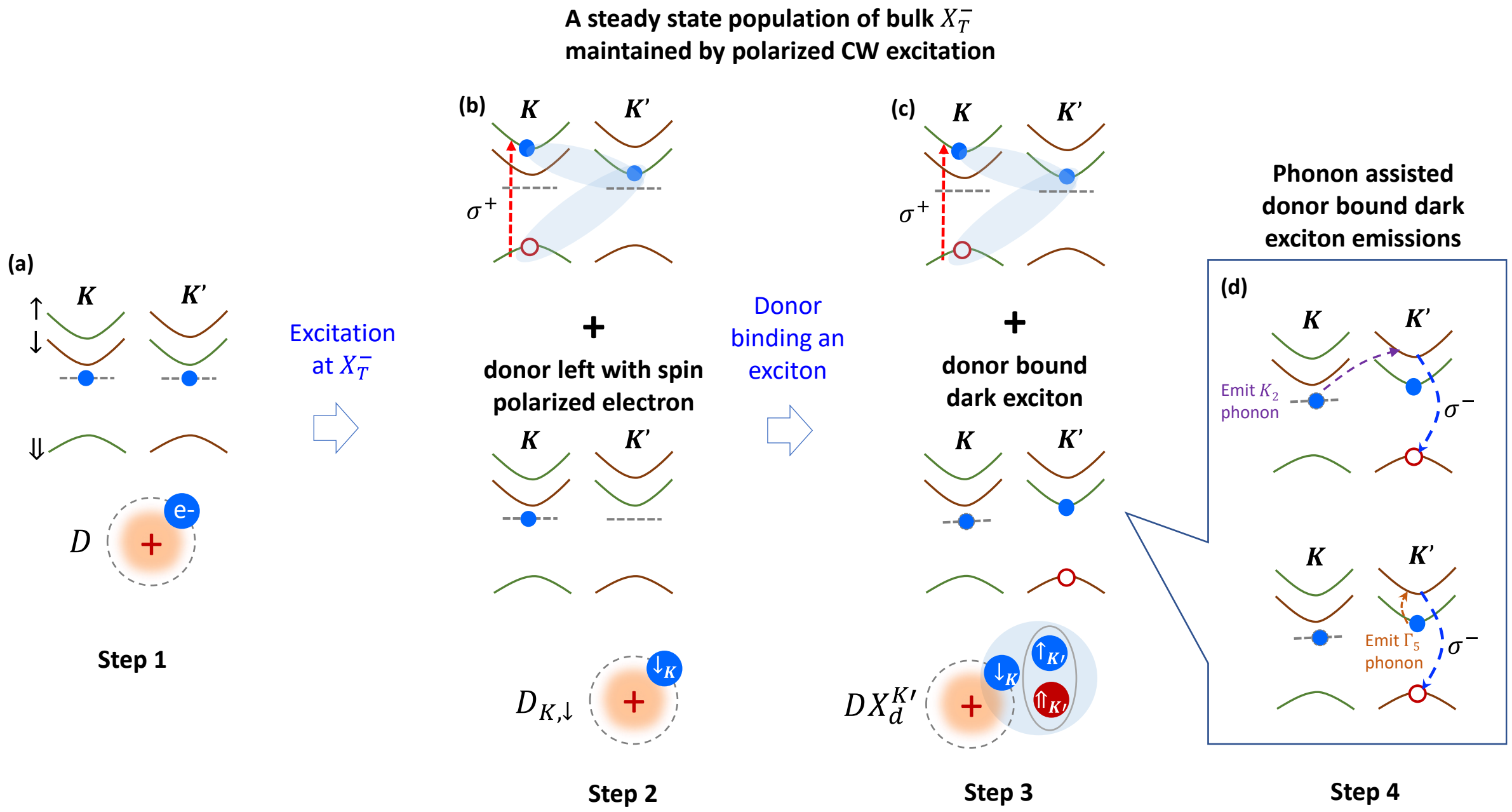

**Supplementary Figure 7. Trion excitation dependent donor bound exciton photoluminescence.** This is an expanded flow diagram of Fig. 5 in the main text and focuses on intervalley trion excitation with $\sigma^+$ polarized light. **a**, step 1 depicts unpolarized donor electrons without optical pumping. **b**, step 2 shows donor spin state initialization by optical pumping of intervalley trion. **c**, step 3 illustrates donor bound exciton formation. **d**, step 4 shows donor bound dark exciton emission assisted by intervalley ($K2$) and intravalley ($\Gamma_5$) phonons. See supplementary note 2 for details.

## Supplementary Note 1. Assignment of donor type defect.

Here, we present collective experimental evidences that the defect resulted in bound excitons is a donor type. Our gate dependent PL shows that defect bound excitons only appear when the monolayer is electron doped. This implies the defect band of interest is near the conduction band minimum (CBM). As shown by Fig 3 in the maintext, we have examined devices made of different crystals obtained by controllable growths. The devices showing defect bound excitons are slightly n doped, while the device made from crystals without defect bound excitons are slightly p doped. All devices are made by the same fabrication procedure. These facts suggest that the initial n doping is unlikely from trapped impurities between monolayer $WSe_2$ and hBN introduced during fabrications. STM study shows that only the devices with bright defects (see Figs. 3a &d) produced defect bound excitons. We further performed scanning tunneling spectroscopy (STS) measurements on these defects. Supplementary Figures 6a and 6b show the STM topographic images of the bright and dark defects, as identified in Fig. 3 in the main text. Supplementary Figures 6c and d show the corresponding differential conductance curves. Clearly, for the crystal producing defect bound excitons, we observe a defect band below CBM (Supplementary Fig. 6c), with binding energy of about 110 meV. There are no other bands observed, ruling out the complicated charge levels within the gap. We further obtained STS on the $WSe_2$ surface close (red) and far (blue) from the donor defect (Supplementary Fig. 6f). Careful analysis of the band edges of the valence and conduction bands on the different locations show that there is an energy offset between them, likely due to band bending. As seen in the figure, the offset between the two curves is about 30 meV, and corresponds to downward band bending near the defect. This supports the donor nature of the defect with positive charging. The defect binding energy in monolayer will be larger than 110 meV, but should be in the same order of magnitude and thus comparable to exciton binding energy in monolayers.[17-22] This supports the shallow potential of the donor. In addition, the devices host defect bound excitons are slightly n doped at zero gate voltage, evident by the negative gate voltage for the charge neutrality point in gate dependent PL, as illustrated in Fig.1c and Fig. 3d in the maintext. This slightly intrinsic n doping is also consistent with the donor nature of the defect. On the other hand, the devices made of crystal in Fig. 3c, which does not have defect bound excitons, are slightly p doped (see gate dependent PL in Fig. 3f in the maintext). This is consistent with the observed defect band near the valence band maximum in Fig. S6d.

**Supplementary Note 2. Trion excitation dependent donor bound exciton photoluminescence.** Here we will further clarify Fig. 5 in the maintext. For simplicity, we only focus on the excitation of intervalley trion with $\sigma^+$ polarized optical excitation (Supplementary Fig. 7).

*Step 1: without optical excitation.* Supplementary Figure 7a depicts the valley-spin-coupled band edges and unpolarized donor bound electrons.

*Step 2: Donor spin state initialization*. As illustrated in the top panel Supplementary Fig. 7b, resonant $\sigma^+$ polarized excitation creates $+K$ valley polarized intervalley trion $X_T^-(+K)$. In this continuous wave excitation experiment, maintaining a steady-state population of $X_T^-(+K)$ "uses up" spin up species of the donor electrons. This optical pumping process then leaves the donor with spin down electrons (denoted as $D_{K,\downarrow}$), as shown in the lower panel of Supplementary Fig. 7b.

*Step 3: Donor bound exciton formation*. The bottom panel of Supplementary Fig. 7c illustrates that the spin polarized donor $D_{K,\downarrow}$ then selectively binds to a dark exciton in the $K'$ valley to form $DX_d^{K'}$, because of the Pauli exclusion between the electrons. Note that a steady state population of the $X_T^-(+K)$ is maintained in this whole process due to $\sigma^+$ polarized continuous wave excitation (top panel Supplementary Fig. 7c). Dark exciton is the ground state and its long lifetime makes possible efficient formation of $DX_d$.

*Step 4: Donor bound exciton light emission.* $DX_d^{K'}$ can emit light via two process. One is via defect mediated direct electron-hole recombination in the K` valley. The second is via phonon assisted stokes emission (Supplementary Fig. 7d). The top panel of Supplementary Fig. 7d shows intervalley spin-conserved electron scattering assisted by $K_2$ phonon, which give rises to S2 peak via an intermediate bright trion state. The bottom panel shows the intravalley spin-flip electron scattering assisted by $\Gamma_5$ phonon for S3 peak.

## Supplementary References